\RequirePackage{filecontents}

\RequirePackage{snapshot}
\documentclass[reprint,aps,pra,floatfix,longbibliography,secnumarabic]{preprint}

\renewcommand\documentclass[2][]{}
\documentclass[reprint,aps,pra,floatfix,longbibliography]{revtex4-1}

\makeatletter
\def\@bibdataout@aps{%
 \immediate\write\@bibdataout{%
  @CONTROL{%
   apsrev41Control%
   \longbibliography@sw{%
    ,author="08",editor="1",pages="0",title="0",year="0"%
   }{%
    ,author="08",editor="1",pages="0",title="",year="1"%
   }%
  }%
 }%
 \if@filesw
  \immediate\write\@auxout{\string\citation{apsrev41Control}}%
 \fi
}%
\makeatother

\usepackage[T1]{fontenc}
\usepackage{amsmath}
\usepackage{txfonts}
\usepackage{tgtermes}
\usepackage[altg]{qtxmath}

\usepackage{microtype}
\usepackage{graphicx}
\usepackage{braket}
\graphicspath{{figures/}} \usepackage[breaklinks=true,
pdfauthor={Nicolas Teeny, Heiko Bauke and Christoph H. Keitel},
pdftitle={Salecker-Wigner-Peres quantum clock applied to strong-field tunnel ionization}]{hyperref}

\newcommand*{\D}{\mathrm{d}}

\newcommand*{\I}{\mathrm{i}}
\newcommand*{\op}[1]{\hat{#1}}

\allowdisplaybreaks

\begin{document}

\title{Salecker-Wigner-Peres quantum clock applied to strong-field
  tunnel ionization}

\author{Nicolas Teeny}
\email{nicolas.teeny@mpi-hd.mpg.de}
\affiliation{Max-Planck-Institut f{\"u}r Kernphysik, Saupfercheckweg~1, 69117~Heidelberg, Germany}
\author{Christoph H. Keitel}
\affiliation{Max-Planck-Institut f{\"u}r Kernphysik, Saupfercheckweg~1, 69117~Heidelberg, Germany}
\author{Heiko Bauke}
\email{heiko.bauke@mpi-hd.mpg.de}
\affiliation{Max-Planck-Institut f{\"u}r Kernphysik, Saupfercheckweg~1, 69117~Heidelberg, Germany}

\begin{abstract}
  The Salecker-Wigner-Peres quantum-clock approach is applied in order
  to determine the tunneling time of an electron in strong-field tunnel
  ionization via a time-dependent electric field.  Our results show that
  the ionization of the electron takes a nonvanishing period of time.
  This tunneling time is of the order of the Keldysh time but strictly
  larger than the Keldysh time.  Comparing the quantum-clock tunneling
  time to the mean tunneling time as obtained by the virtual-detector
  approach,
  one finds that these two complementary methods give very similar
  results.  Due to the asymmetric distribution of the tunneling time,
  there is a nonnegligible discrepancy between the mean tunneling time
  and the most probable tunneling time.
\end{abstract}

\maketitle

\enlargethispage{2pt}

\section{Introduction}
\label{sec:intro}

Tunneling and tunnel ionization are fundamental processes in quantum
mechanics, which are not only of theoretical interest but are also the
foundation of some technical applications.  A respectable stock of
scientific works has been devoted to this topic.  Nevertheless, some
aspects of tunneling are discussed controversially till today, in
particular the temporal development of the tunneling dynamics and the
time span that is required to cross the tunneling barrier.  The issue of
tunneling times was first considered by \mbox{MacColl} in
Ref.~\cite{MacColl:1932:Note_on} for a free particle tunneling through a
static square potential.  In this case, a particle approaches from far
away a potential wall higher than its initial kinetic energy.  Since
\mbox{MacColl's} pioneering work many approaches and methods have been
proposed to define a tunneling time
\cite{Hauge:Stvneng:1989:Tunneling_times,Landauer:Martin:1994:Barrier_interaction}
for this physical situation, which can be classified into three
categories.

A possible and very intuitive approach is to determine the traversal
time by following the center of gravity of the transmitted wave packet
\cite{Martin:Landauer:1992:Time_delay}.  The associated time, however,
has little physical significance as argued in
Ref.~\cite{Landauer:Martin:1994:Barrier_interaction}.  The second class
of approaches constructs a set of dynamical paths and determines how
much time each path spends under the tunneling barrier.  Then, one can
define the most probable time spent under the barrier, corresponding to
the most probable path, or an average time spent under the barrier by
taking a weighted average over all paths.  Among others, this approach
is realized by 
the Bohm method as described in
Refs.~\cite{Leavens:Aers:1996:Bohm_Trajectories,Leavens:1995:Bohm_trajectory}
and references therein, the Feynman path integrals method as applied in
Refs.~\cite{Sokolovski:Baskin:1987:Traversal_time,%
  Fertig:1990:Traversal-Time_Distribution,%
  Leavens:1995:Bohm_trajectory}, and finally the Wigner distribution
paths method as studied in
Refs.~\cite{Jensen:Buot:1989:Numerical_calculation,%
  Balazs:Voros:1990:Wigners_function,%
  Marinov:Segev:1996:Quantum_tunneling}.  Time is not only a coordinate
of the universal space-time background where physical processes take
place.  Time can also be introduced as a dynamical variable of physical
systems that clock a certain process \cite{Hilgevoord:2002:Time_in},
which leads us to the third category of approaches to define a tunneling
time.  In the so-called quantum-clock approach, an additional physical
system is coupled to the system which undergoes the tunneling dynamics
\cite{Salecker:Wigner:1958:Quantum_Limitations,%
  Baz:1967:Lifetime,%
  Rybachenko:1967:Time,%
  Peres:1980:Measurement_of,%
  Buttiker:1983:Larmor_precession,%
  Martin:Bianchi:1992:On_theory}.  Then either a dynamical variable of
the coupled system acts as a clock or the accessory system has an
explicit time dependence with a given time scale, which provides a
reference for time measurements
\cite{Buttiker:Landauer:1982:Traversal_Time,Buttiker:Landauer:1985:Traversal_Time}.
Depending on how temporal quantities are extracted from the clock
system, the quantum-clock approach gives rise to definitions of various
times which characterize the tunneling dynamics, e.\,g., the dwell time,
the tunneling (traversal) time, or the reflection time.

Early works on tunnling times mainly focus on tunneling prosesses where
an asymptotically free particle approaches a static potential barrier,
tunnels though the barrier, and finally becomes free again.  Currently,
the issue of tunnel ionization times got into the focus of scientific
research due to the progress in experimental atomic physics, which
allows one to probe strong-field ionization dynamics at attosecond time
scales
\cite{Eckle:Smolarski:etal:2008:Attosecond_angular,Eckle:Pfeiffer:etal:2008:Attosecond_Ionization}.
These experiments are often referred to as attoclock experiments.
Tunneling of initially free particles is quite different from tunnel
ionization
\cite{Ban:Sherman:etal:2010:Time_scales,Orlando:McDonald:etal:2014:Tunnelling_time}
in atomic physics.  Here, the particle is initially bound by a binding
potential and then excited by a time-dependent electric field pulse
which induces the tunneling dynamics.  Because the typical time scale of
the driving electric field is large compared to the typical tunneling
time, the electron's wave function has time to adapt to the changing
potential barrier.  Since these attoclock experiments have been
performed, many renewed efforts have been directed toward defining a
tunnel ionization time because a consensus on a suitable theoretical
definition of tunneling time and the interpretation of experimental
results is still lacking \cite{Keldysh:1965:Ionization,%
  Eckle:Smolarski:etal:2008:Attosecond_angular,%
  Eckle:Pfeiffer:etal:2008:Attosecond_Ionization,%
  Ban:Sherman:etal:2010:Time_scales,%
  Lein:2011:Streaking_analysis,%
  Yakaboylu:Klaiber:etal:2013:Relativistic_features,%
  McDonald:Orlando:etal:2013:Tunnel_Ionization,%
  Klaiber:Yakaboylu:etal:2013:Under-the-Barrier_Dynamics,%
  Zhao:Lein:2013:Determination_of,%
  Kaushal:Smirnova:2013:Nonadiabatic_Coulomb,%
  Landsman:Weger:etal:2014:Ultrafast_resolution,%
  Orlando:McDonald:etal:2014:Tunnelling_time,%
  Yakaboylu:Klaiber:etal:2014:Wigner_time,%
  Maquet:Caillat:etal:2014:Attosecond_delays,%
  Landsman:Keller:2015:Attosecond_science,%
  Torlina:Morales:etal:2015:Interpreting_attoclock,%
  Klaiber:Hatsagortsyan:etal:2015:Tunneling_Dynamics,%
  Teeny:etal:2016:Ionization_Time,%
  Teeny:2016:virt_detect}.

For studying tunneling times in tunnel ionization theoretically, many
approaches can be adopted from tunneling of initially free particles.
For example, the Wigner time approach
\cite{Sokolovski:Baskin:1987:Traversal_time,Wigner:1955:Lower_Limit,Smith:1960:Lifetime_Matrix}
was applied to tunnel ionization in the adiabatic limit in
Refs.~\cite{Yakaboylu:Klaiber:etal:2014:Wigner_time,Yakaboylu:Klaiber:etal:2013:Relativistic_features}.
The adiabatic limit corresponds to a parameter regime where the time
scale of the tunneling dynamics is short compared to the time scale of
the variation of the electric field.  Calculating the complex
transmission amplitude as a function of the barrier height and the
electron energy, various theoretical definitions of tunneling times can
be introduced, often referred to as B{\"u}ttiker-Landauer time,
Pollack-Miller time, Eisenbud-Wigner time, and Lamor time.  These have
been compared to experimental results in
Refs.~\cite{Landsman:Weger:etal:2014:Ultrafast_resolution,Zimmermann:Mishra:etal:2016:Tunneling_Time}.
The interpretation of the attoclock measurements is, however, not
trivial and depends on the employed theoretical model of the ionization
dynamics \cite{Torlina:Morales:etal:2015:Interpreting_attoclock}.  It is
a commonly applied assumption that ionization happens at the instant of
the electric field maximum.  It has, however, been shown recently by
applying the virtual-detector approach to strong-field tunnel ionization
in Refs.~\cite{Teeny:etal:2016:Ionization_Time,Teeny:2016:virt_detect}
that the moment when the electron leaves the tunneling barrier does not
necessarily coincide with the moment of electric field maximum.

Except the virtual-detector approach, none of the above mentioned
approaches has been applied to tunnel ionization in a time-dependent
potential, i.\,e., taking into account the continuous increase and decay
of the external driving electric field as it is the situation in an
experimental setting.  Often the external field is treated as static
\cite{Yakaboylu:Klaiber:etal:2014:Wigner_time} or switched on
instantaneously
\cite{Ban:Sherman:etal:2010:Time_scales,Orlando:McDonald:etal:2014:Tunnelling_time}.
As emphasized in Ref.~\cite{Orlando:McDonald:etal:2014:Tunnelling_time},
tunneling in a continuously evolving potential is very different from
the sudden turn-on case.  In particular, in a slowly varying electric
field there is no natural reference point in time which defines when
tunneling begins.  Furthermore, the quantum state at the onset of
tunneling is no longer the ground state of the unperturbed binding
potential.  When tunneling sets in, the wave function has already
evolved in the time-dependent potential.

In this work, we apply the Salecker-Wigner-Peres quantum clock to
strong-field tunnel ionization taking into account the continuous
evolution of the driving electric field to determine the the dwell time,
the traversal time, and the reflection time.  The obtained traversal
time is compared to the mean tunneling time as calculated by the
recently developed virtual-detector method
\cite{Teeny:2016:virt_detect,Teeny:etal:2016:Ionization_Time}.  The
article is organized as follows: In Sec.~\ref{sec::system}, the
considered system is described.  The Salecker-Wigner-Peres quantum clock
is reviewed in Sec.~\ref{sec::clock}, before we explain how to apply
this approach to determine the tunneling time of tunnel ionization.  Our
main results are presented and interpreted in Sec.~\ref{sec::results}.
In Sec.~\ref{sec:Conclusions} finally, we summarize our main results.

\section{Tunnel ionization from a two-dimensional Coulomb potential}
\label{sec::system}

In the following, we will study tunnel ionization from a two-dimensional
Coulomb potential induced via a driving homogenous electric field.  This
two-dimensional system resembles tunnel ionization from hydrogen-like
ions while keeping the computational demands small.  In the
long-wavelength limit, i.\,e., when the dipole approximation is
applicable, the three-dimensional Coulomb potential with an external
electric field has rotational symmetry around the electric field
direction, which makes this system quasi two-dimensional.

Choosing the coordinate system such that the linearly polarized external
electric field with the amplitude $\mathcal{E}(t)$ points into the $x$
direction, the Schr{\"o}dinger equation for the Coulomb problem with the
Hamiltonian $\op{H}_E$ reads
\begin{multline}
  \label{eq:Schrodinger_2d}
  \I\hbar\frac{\partial\Psi(x, y, t)}{\partial t} = 
  \op{H}_E\Psi(x,y,t)= \\
  \left(
    -\frac{\hbar^2}{2m}\left(
      \frac{\partial^2}{\partial x^2}+
      \frac{\partial^2}{\partial y^2}
    \right)
    -\frac{Ze^2}{4\pi\varepsilon_0\sqrt{x^2+y^2}} - ex\mathcal{E}(t)
  \right)\Psi(x, y, t)\,.
\end{multline}
Here $m$, $e$, $Z$, and $\varepsilon_0$ denote the electron's mass, the
elementary charge, the atom's atomic number, and the vacuum
permittivity, respectively.  Applying an electric field pulse with a
unique maximum allows us to study the ionization dynamics without
undesirable artifacts, i.\,e., to avoid multiple ionization and
rescattering.  Therefore, we employ a Gaussian pulse, i.\,e., the
electric field is given by
\begin{equation}
  \label{eq:E_t}
  \mathcal{E}(t)=\mathcal{E}_0\exp\left(
    -\tfrac{\omega_\mathcal{E}^2(t-t_0)^2}{2}
  \right)\,.
\end{equation}
The time $t_0$ denotes the instant of maximal electric field
$\mathcal{E}_0$ and $\tau_{\mathcal{E}} = \sqrt{2}/\omega_\mathcal{E}$
is the time scale of the raise and decay of the electric field.  Note
that at $t\approx t_0$ the Gaussian pulse \eqref{eq:E_t} corresponds
approximately to a sinusoidal pulse with
$\mathcal{E}(t)=\mathcal{E}_0\cos(\omega_\mathcal{E} t)$.

\section{Salecker-Wigner-Peres quantum clock}
\label{sec::clock}

\subsection{Fundamentals}
\label{sec::clock:fundamentals}

Salecker, Wigner, and Peres
\cite{Salecker:Wigner:1958:Quantum_Limitations,Peres:1980:Measurement_of}
introduced a quantum system that can serve as a clock, the so-called
Salecker-Wigner-Peres quantum clock. This clock system is characterized
by the Hamiltonian $\op{H}_c=\omega\op{J}$, where
$\op{J}=-\I\hbar\frac{\partial}{\partial\theta}$ denotes an angular
momentum operator for the angular coordinate $\theta$ and $\omega$ is an
angular frequency.  Obviously, the operators $\op{H}_c$ and $\op{J}$
share a common set of eigen-functions.  Restricting the angular variable
$\theta$ to $\theta\in[0,2\pi)$ and imposing periodic boundary
conditions, the clock Hamiltonian $\op{H}_c$ and the operator $\op{J}$
possess the equidistant discrete spectra $n\hbar\omega$ and $n\hbar$,
respectively, with integer $n$.  The corresponding normalized
eigen-functions will be denoted by $\ket{J_n}$ in the following.

For the subspace that is spanned by the orthonormal functions
$\ket{J_{-j}}$, \dots, $\ket{J_{j}}$ the following new basis set can be
introduced:
\begin{equation}
  \ket{V_k} = 
  \frac{1}{\sqrt{N}} \sum_{n=-j}^j\exp\left(
    -\tfrac{2\pi\I kn}{N}
  \right)\ket{J_n} \,,
\end{equation}
with $N=2j+1$ and $k=0,1,\dots,N-1$.  The $N$ states $\ket{V_k}$ are
called clock states.  Under the evolution of the clock Hamiltonian, an
initial clock state cycles successively through all clock states.  A
short calculations shows for $\Delta t=2\pi/(N\omega)$
\begin{multline}
  \label{eq:H_c_V_k}
  \exp\left(
    -\tfrac{\I\op{H}_c\,\Delta t}{\hbar}
  \right) \ket{V_k} = 
  \frac{1}{\sqrt{N}} \sum_{n=-j}^j
  \exp\left(
    -\tfrac{2\pi\I n}{N}
  \right)
  \exp\left(
    -\tfrac{2\pi\I kn}{N}
  \right)\ket{J_n} \\ =  
  \ket{V_{(k+1)\bmod N}} \,.
\end{multline}
This means, after preparing $\ket{V_0}$ as the clock's initial state,
the clock's quantum state passes trough $\ket{V_0}$, $\ket{V_1}$,
$\ket{V_2}$, and so on at times $t=0$, $t=\Delta t$, $t=2\,\Delta t$,
and so on until $\ket{V_{N-1}}$ passes into $\ket{V_0}$ after a further
time step of $\Delta t$.

\begin{figure}
  \centering
  \includegraphics[scale=0.45]{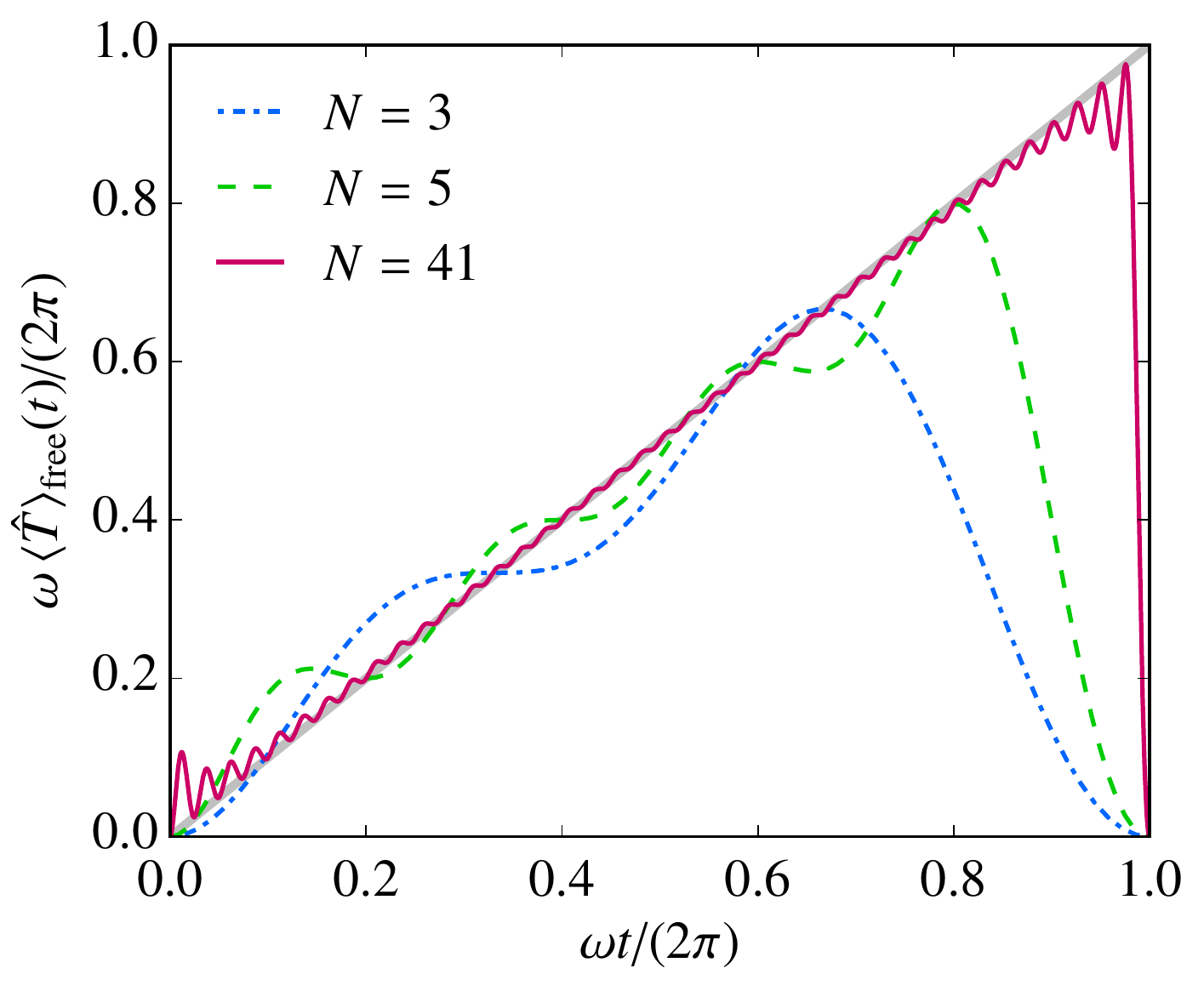}
  \caption{Expectation value of the clock operator
    $\langle\hat{T}\rangle=f(t)$ as a function of time $t$ for freely
    running clocks with different numbers of clock states $N$ and
    $\ket{V_0}$ as the initial quantum state.  As a reference, the solid
    gray line indicates linear passage of time.}
  \label{fig:calibration}
\end{figure}

Introducing the set of projection operators $\op{P}_k$ which fulfill
\begin{equation}
  \op{P}_k \ket{V_l} =  
  \big(\ket{V_k} \!\bra{V_k} \big)\ket{V_l} = 
  \delta_{k,l}\ket{V_k} 
\end{equation}
for $k,l=0,1,\dots,N-1$, one can define the clock operator
\begin{equation}
  \label{eq:time_operator}
  \op{T} = \sum_{k=0}^{N-1} \Delta t\, k\op{P}_k \,.
\end{equation}
The time-dependent expectation value of the
operator~\eqref{eq:time_operator} for a free-running clock with
$\ket{V_0}$ as its initial state
\begin{equation}
  \label{eq:free_clock}
  \braket{\op{T}}_\mathrm{free}(t) = 
  \Braket{ V_0 | 
    \exp\left(
      \tfrac{\I\op{H}_ct}{\hbar}
    \right)\op{T} \exp\left(
      -\tfrac{\I\op{H}_ct}{\hbar}
    \right) |
    V_0} 
\end{equation}
equals at times $t=n\,\Delta t$ 
\begin{equation}
  \braket{\op{T}(n\,\Delta t)}_\mathrm{free} =
  \Braket{V_{n \bmod N}|\op{T}|V_{n \bmod N}} = 
  \Delta t\,(n \bmod N) \,,
\end{equation}
as a consequences of Eq.~\eqref{eq:H_c_V_k}.  The expectation value
\eqref{eq:free_clock} is shown in Fig.~\ref{fig:calibration} as a
function of time $t$ and different numbers of clock states $N$.  As seen
in this figure, for noninterger multiples of $\Delta t$, the expectation
value of the clock operator deviates from the laboratory time~$t$.
Thus, the quantity $\Delta t$ characterizes the resolution of the
quantum clock.  In particular at $t\gtrapprox0$ and
$t\lessapprox 2\pi/\omega=N\,\Delta t$, the periodicity of the clock
operator induces large oscillations near the clock's discontinuity at
$t=N\,\Delta t$ similar to the Gibbs-Wilbraham
phenomenon~\cite{Hewitt:Hewitt:1979:Gibbs-Wilbraham_phenomenon}.

\subsection{Coupling the Salecker-Wigner-Peres quantum clock
  to~a~dynamical system}
\label{sec::clock:coupling}

In order to measure the duration of a dynamical process in some quantum
system with the Hamiltonian $\op H$, the Salecker-Wigner-Peres quantum
clock has to be coupled to the system of interest.  The Hilbert space of
the combined system with the Hamiltonian $\op{H}_T$ becomes the tensor
product of the Hilbert space of $\op{H}$ with the state vector
$\ket{\Psi}$ and of the clock Hamiltonian $\op{H}_c$ with the state
vector $\ket{\Psi_c}$.  Thus, the quantum state of the combined system
becomes $\ket{\Phi}=\ket{\Psi}\otimes\ket{\Psi_c}$, which may be
represented by an \mbox{$N$-component} wave function in case of a
\mbox{$N$-level} clock system.  To give a specific example, let us
consider the time of flight of a free particle of mass $m$
\cite{Peres:1980:Measurement_of,Leavens:1993:Application_of} in one
dimension.  In this case, the one-dimensional Hamiltonian $\op{H}$
coupled to a clock Hamiltonian $\op{H}_c$ with $N=2j+1$ states is given
by \cite{Peres:1980:Measurement_of}
\begin{multline}
  \label{eq:time-flight}
  \op{H}_T = \op{H} + \op{P}([0, \ell])\op{H}_c = \\
  -\frac{\hbar^2}{2m}\frac{\D^2}{\D x^2} + 
  \op{P}([0, \ell]) 
  \hbar\omega
  \begin{pmatrix}
    j \\
    & j-1 \\
    & & \ddots \\
    & & & -j
  \end{pmatrix} \,.
\end{multline} 
Here, $\op{H}_c$ is represented in its eigen-basis and
$\op{P}([0, \ell])$ denotes the projection operator on $x \in[0, \ell]$,
which is the region of the time-of-flight measurement.  This operator
yields one for $x \in[0, \ell]$ and zero otherwise.  Using the
Hamiltonian~\eqref{eq:time-flight}, one is able to measure how much time
the free particle spends in the region $x \in[0, \ell]$ because the
coupling of the wave function to the clock Hamiltonian $\op{H}_c$ is
restricted to this range.  To determine the time of flight, the state
\mbox{$\ket{\Psi_0}\otimes\ket{V_0}$} is prepared as the initial
condition, such that $\ket{\Psi_0}$ represents a particle far away from
$[0, \ell]$.  The the time operator~\eqref{eq:time_operator} is applied
to the evolving quantum state $\ket{\Phi}$ of the combined system.

Figure~\ref{fig:calibration} shows that for large $N$, the expectation
value of the clock operator follows closely the laboratory time~$t$, in
particular at $t\approx \pi/\omega$.  Nevertheless, it is not desirable
to choose a large number of clock states $N$ because the extraneous
disturbance of the quantum system due to the clock Hamiltonian is also
proportional to $N$.  In order not to disturb the quantum system under
investigation by the clock Hamiltonian, we set $N=3$ in the following.
For $N=3$, the clock operator's expectation value grows monotonously
with time $t$, at least for $t\in[0, 2\,\Delta t]$.  Thus, we can invert the
function $\braket{\op{T}}_\mathrm{free}(t)$, which yields the
expectation value of the operator~\eqref{eq:time_operator} of a free
running clock, to calibrate the quantum clock
\cite{Leavens:1993:Application_of}.  In this way, the laboratory time
$t$ can be inferred from the clock operator's expectation value to much
higher accuracy than given by $\Delta t$.

\subsection{Determining tunneling times in tunnel ionization}
\label{sec::clock:measurement}

\begin{figure}
  \centering
  \includegraphics[scale=0.45]{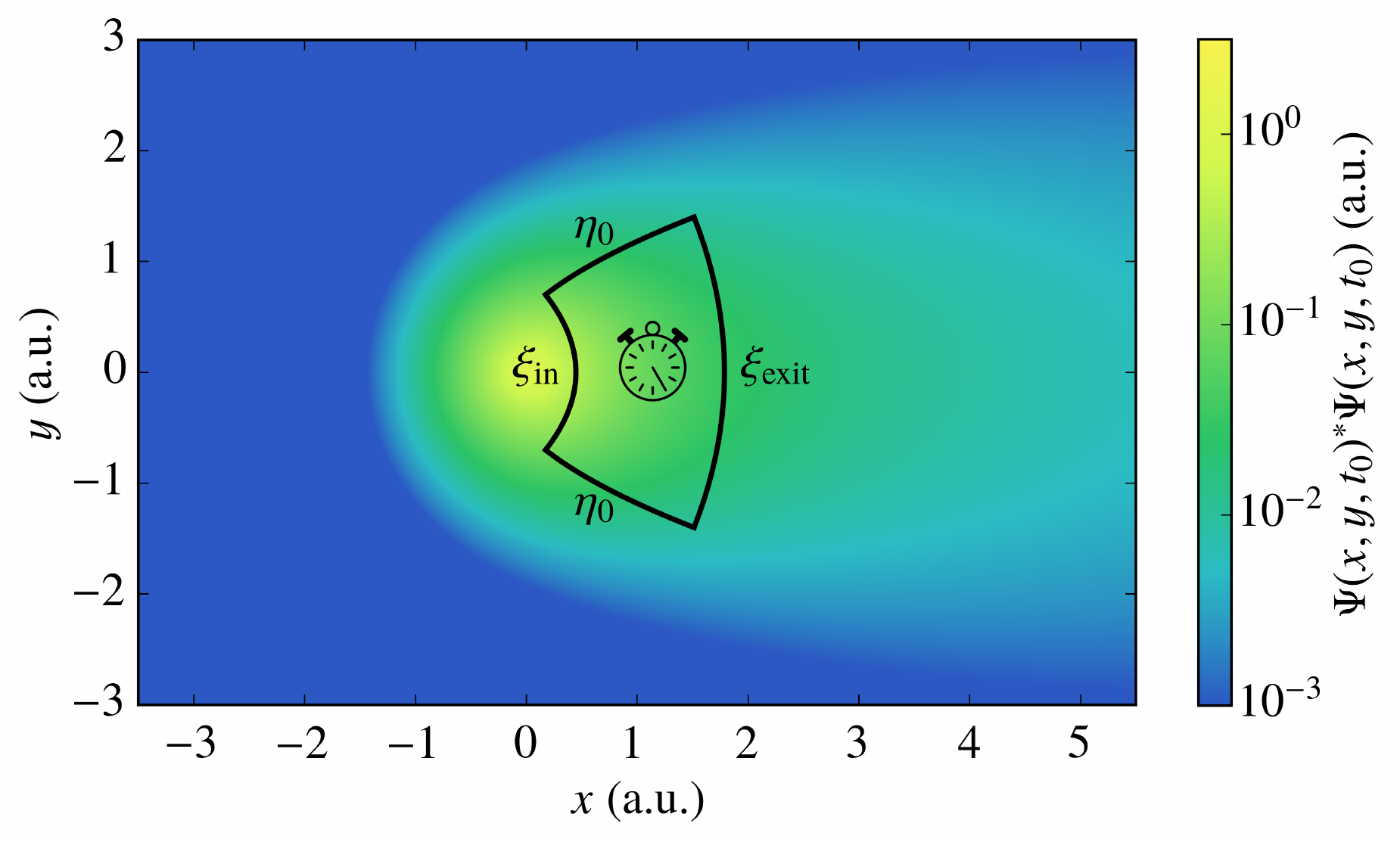}
  \caption{Schematic of tunneling time determination in strong-field
    ionization via the Salecker-Wigner-Peres quantum clock.  The
    coupling of the clock's Hamiltonian to electron's Hamiltonian is
    established in the tunneling barrier region $\mathcal{B}$ between
    the lines $\eta=\eta_0$, $\xi=\xi_\mathrm{in}$ and
    $\xi=\xi_\mathrm{exit}$ indicated by the black solid lines; see
    Ref.~\cite{Teeny:2016:virt_detect}. The wave function's probability
    density $|\Psi(x, y, t)|^2$ at the instant of electric field
    maximum, i.\,e., at $t=t_0$, is represented by colors for
    $\mathcal{E}_0=1.1\,\mathrm{a.u.}$, $Z=1$, and $\gamma=0.25$.}
  \label{fig:barrier}
\end{figure}

In tunnel ionization of an electron initially bound to a Coulomb
potential, the space can be divided in to a classically allowed region
and a classically forbidden region, which also contains the tunneling
region.  The shape of the tunneling region is formed by bending the
atomic binding potential via the applied external electric field and the
electron's ground-state energy corrected by Stark-shift effects; see
Fig.~\ref{fig:barrier} and Ref.~\cite{Teeny:2016:virt_detect} for
details.  Introducing the parabolic coordinates $\xi$ and $\eta$ via
\begin{align}
  x & = \frac{\xi - \eta}{2} \,, &
  y & = \sqrt{\xi\eta} \,,
\end{align}
the tunneling region is confined by lines of constant $\xi$ and~$\eta$.
In order to measure the tunneling ionization time by the quantum clock,
the coupling between the quantum clock and the electron is established
in the tunneling barrier region $\mathcal{B}$ via the projection
operator $\op{P}(\mathcal{B})$.  Thus, the corresponding Hamiltonian is
given by
\begin{equation}
  \label{eq:coupling-elec-clock}
  \op{H}_T = \op{H}_E + \op{P}(\mathcal{B})\op{H_c} \,,
\end{equation}
where $\op H_E$ is the electron Hamiltonian as defined in
Eq.~\eqref{eq:Schrodinger_2d} and $\op{H_c}$ denotes an $N$-level clock
Hamiltonian with $N=2j+1=3$.  The combined wave function $\ket{\Phi(t)}$
of the electron and the clock is initially given by
$\ket{\Phi(0)}=\ket{\Psi_0}\otimes\ket{V_0}$, where $\ket{\Psi_0}$
denotes the ground state of the two-dimensional Coulomb potential.

In general, the tunneling region is time-dependent due to the
time-dependent electric field \eqref{eq:E_t}.  Nevertheless, we consider
a time-independent tunneling region which is given for the peak of the
applied electric field at time $t=t_0$.  Coupling the quantum clock to
the electron in this fixed tunneling region is justified because for
static fields the tunneling probability is maximal for maximal electric
fields and it is exponentially suppressed for lower fields.

The projection operator $\op{P}(\mathcal{B})$ in
Eq.~\eqref{eq:coupling-elec-clock} is designed to advance the quantum
clock only when the electron is in the tunneling region.  However, the
determination of the tunneling time via the quantum clock is complicated
by the fact that there is always a small portion of the electron wave
packet in the tunneling region even at a vanishing electric field,
i.\,e., before the tunneling dynamics starts.  Moreover, only a part of
the probability density which enters the tunneling region will
eventually escape the tunneling region and become free.  The other part
gets just reflected under the tunneling barrier.  Since the quantum
clock cannot differentiate between the reflected and the tunneled part
of the electron wave packet, it determines the time spent in the
tunneling barrier region $\mathcal{B}$ irrespective whether the electron
is eventually reflected or transmitted.  This time is known as the dwell
time (or sojourn time) $\tilde{\tau}_D$ and was discriminated clearly
from other concepts of tunneling times for the first time by B{\"u}ttiker
\cite{Buttiker:1983:Larmor_precession}.  The following alternative
expression
\begin{equation}
  \label{eq:t-dwell}
  \tau_D' = 
  \int_{0}^{\infty}\int_{\mathcal{B}}|\Psi(x,y,t)|^2\,\D x\,\D y\,\D t
\end{equation}  
for the dwell time can be derived within conventional
\cite{Sokolovski:Baskin:1987:Traversal_time} as well as within Bohm's
\cite{Leavens:1990:Transmission_reflection} interpretations of quantum
mechanics and is commonly accepted by now
\cite{Olkhovsky:Recami:1992:Recent_developments}.

The tunneling time can be derived from the dwell time or the asymptotic
expectation value of the clock operator \eqref{eq:time_operator} applied
to the wave function $\ket{\Phi}$, respectively, via splitting this time
into a weighted sum of a tunneling time $\tilde{\tau}_T$ and a
reflection time $\tilde{\tau}_R$ as shown in the following.  After the
tunneling dynamics has finished, the quantum state $\ket{\Phi}$ can be
separated into a bound part $\ket{\Phi_\mathrm{bound}}$ and a free part
$\ket{\Phi_\mathrm{free}}$ which both occupy two disjunct space regions.
Such regions may be defined via a sphere of sufficiently large radius
around the atomic core, which disjoins both regions.  Thus, the quantum
state $\ket{\Phi}$ after the interaction with the driving electric
field, can be written as a superposition of the two orthonormal states
$\ket{\Phi_\mathrm{bound}}$ and $\ket{\Phi_\mathrm{free}}$ as
\begin{equation}
  \ket{\Phi} = 
  \sqrt{T}\,\ket{\Phi_\mathrm{free}} + \sqrt{R}\,\ket{\Phi_\mathrm{bound}} \,,
\end{equation}
where $T$ denotes the total tunneling probability and $R$ is the
deflection probability with $1=T+R$.  Consequently, the asymptotic
expectation value of the clock operator
\begin{equation}
   \lim_{t\to\infty}\Braket{\Phi(t)|\op T|\Phi(t)} = 
   \Braket{\Phi|\op T|\Phi} = 
   \tilde{\tau}_D 
\end{equation}
may be written as
\begin{equation}
  \label{eq:split_dwell_time}
  \tilde{\tau}_D = T \tilde{\tau}_T + R \tilde{\tau}_R \,,
\end{equation}
where we have introduced the (not calibrated) tunneling time
\begin{subequations}
  \begin{equation}
    \tilde{\tau}_T =
    \Braket{\Phi_\mathrm{free}|\op T|\Phi_\mathrm{free}} 
  \end{equation}
  and the (not calibrated) reflection time
  \begin{equation}
    \tilde{\tau}_R =
    \Braket{\Phi_\mathrm{bound}|\op T|\Phi_\mathrm{bound}} \,.
  \end{equation}
\end{subequations}

The time $\tilde{\tau}_D$ but also $\tilde{\tau}_T$ and $\tilde{\tau}_R$
do not grow proportionally to the laboratory time $t$.  For example, the
physical dwell time is related to $\tilde{\tau}_D$ via
$\tilde{\tau}_D=\braket{\op{T}}_\mathrm{free}(\tau_D)$, as noted in
Sec.~\ref{sec::clock:coupling}.  Thus, the expectation values
$\tilde{\tau}_D$, $\tilde{\tau}_T$, and $\tilde{\tau}_R$ have to be
corrected via inverting $\braket{\op{T}}_\mathrm{free}(t)$ to get the
physical dwell, tunneling, and reflection times
\begin{align}
  \label{eq:tau_D}
  \tau_D & = \overline{\braket{\op{T}}_\mathrm{free}}(\tilde{\tau}_D) \,,\\
  \label{eq:tau_T}
  \tau_T & = \overline{\braket{\op{T}}_\mathrm{free}}(\tilde{\tau}_T) \,,\\
  \label{eq:tau_R}
  \tau_R & = \overline{\braket{\op{T}}_\mathrm{free}}(\tilde{\tau}_R) \,,
\end{align}
where the bar indicates the function's inverse.

Although the splitting \eqref{eq:split_dwell_time} has been employed in
many works \cite{Steinberg:1995:How_Much,Muga:Brouard:etal:1992:Transmission_and}, it has also been criticized
\cite{Landauer:Martin:1994:Barrier_interaction,Leavens:1995:tunneling_time,Goto:Iwamoto:etal:2004:Relationship_between}.
One point of criticism that was put forward is that the dwell time
\eqref{eq:t-dwell} adds up probability density rather than probability
amplitudes and therefore neglects possible interferences between
transmitted and reflected portions of the electron's wave packet.  This
argument, however, does not apply to the clock approach taken here.  The
clock Hamiltonian couples to the wave function, not to the density.
Furthermore, the expectation value \eqref{eq:split_dwell_time} is
calculated when the tunneled and the bound parts of the wave function
are well separated and therefore there is no interference between both
parts.  For the quantum-clock approach it is not required to separate
the wave function into tunneling and reflecting parts under the barrier,
which would be problematic indeed.

\section{Numerical results and interpretation}
\label{sec::results}

\subsection{Dwell time, tunneling time, and reflection time}

\begin{figure}[b]
  \centering
  \includegraphics[scale=0.45]{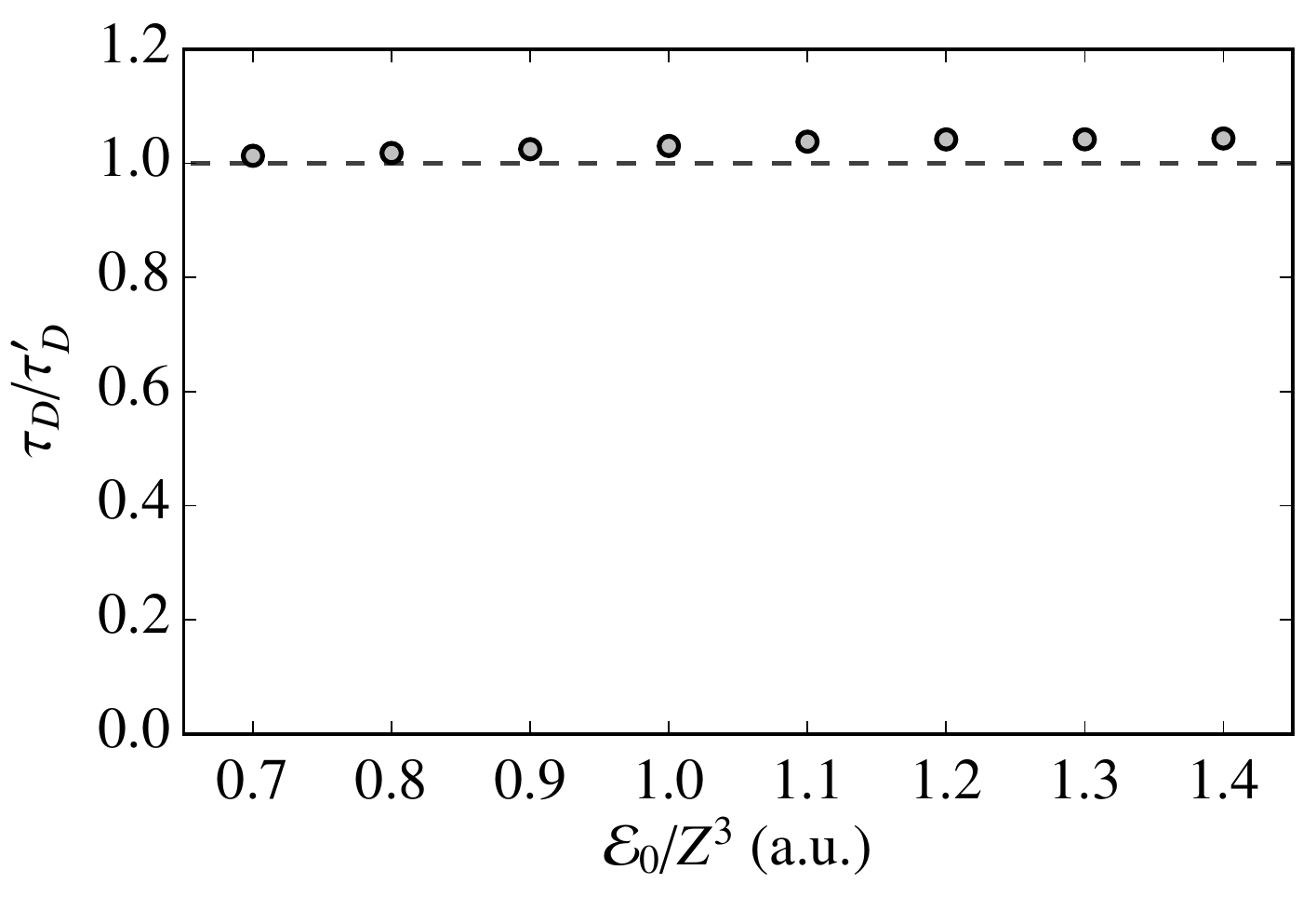}
  \caption{The ratio of the dwell time $\tau_D'$ determined by
    integrating the probability density in the tunneling barrier to the
    (corrected) dwell time $\tau_D$ as given by the quantum clock for
    different electric field strengths $\mathcal{E}_0$ and for a
    constant Keldysh parameter $\gamma=0.25$.  To guide the eye, the
    horizontal dashed line indicates the ratio one, i.\,e., when both
    dwell times exactly agree.}
  \label{fig:dwell}
\end{figure}

After having specified the theoretical foundations of our quantum-clock
approach to tunnel ionization in the previous sections, we can present
the numerical results as obtained by solving the Schr{\"o}dinger equation
with the Hamiltonian~\eqref{eq:coupling-elec-clock} numerically, see the
\hyperref[sec:methods]{Appendix} for details regarding the numerical
methods.  The so-called Keldysh parameter
$\gamma=\omega_\mathcal{E}\sqrt{-2E_Em}/(e\mathcal{E}_0)$
\cite{Keldysh:1965:Ionization} characterizes the ionization process as
dominated by tunneling for $\gamma\ll1$ or by multiphoton ionization for
$\gamma\gg1$.  Here, $E_E$ denotes the ground state binding energy,
which equals $E_E=-2Z^2\,\mathrm{a.u.}$ for the two-dimensional Coulomb
problem \cite{Yang:Guo:etal:1991:Analytic_solution}.  In the following,
the electric field amplitude $\mathcal{E}_0$ and the frequency
$\omega_\mathcal{E}$ are adjusted such that $\gamma=0.25<1$. The clock
parameters $\Delta t=200/Z^2\,\mathrm{a.u.}$ and $N=3$ are chosen such
that $|j\hbar\omega/E_E|=0.0104\ll 1$, i.\,e., the quantum clock
Hamiltonian is coupled weakly to the electron's ave function.  This
ensures that the perturbation of the Coulomb Hamiltonian $\op{H}_E$ by
the quantum clock Hamiltonian $\op{H}_c$ is negligible.  This has been
tested by repeating numerical calculations with
$\Delta t=400/Z^2\,\mathrm{a.u.}$, i.\,e., even weaker coupling, which
yields results that agree with the results obtained for
$\Delta t=200/Z^2\,\mathrm{a.u.}$ up to small numerical discrepancies.
Note that choosing the clock parameter $\Delta t$ arbitrarily large such
that the influence of the clock on the studied system becomes
arbitrarily small leads to numerical difficulties.  With very small
$\Delta t$ also transitions between various clock states $\ket{V_k}$
become small and therefore difficult to resolve numerically.  As
explained in Sec.~\ref{sec::clock:coupling} increasing the number of
states of the clock system does not improve the clock precision, and
thus we choose the smallest nontrivial odd number of states $N=3$.  A
larger number of states $N$ would increase the required numerical effort
without providing any advantage.

\begin{figure}
  \centering
  \includegraphics[scale=0.45]{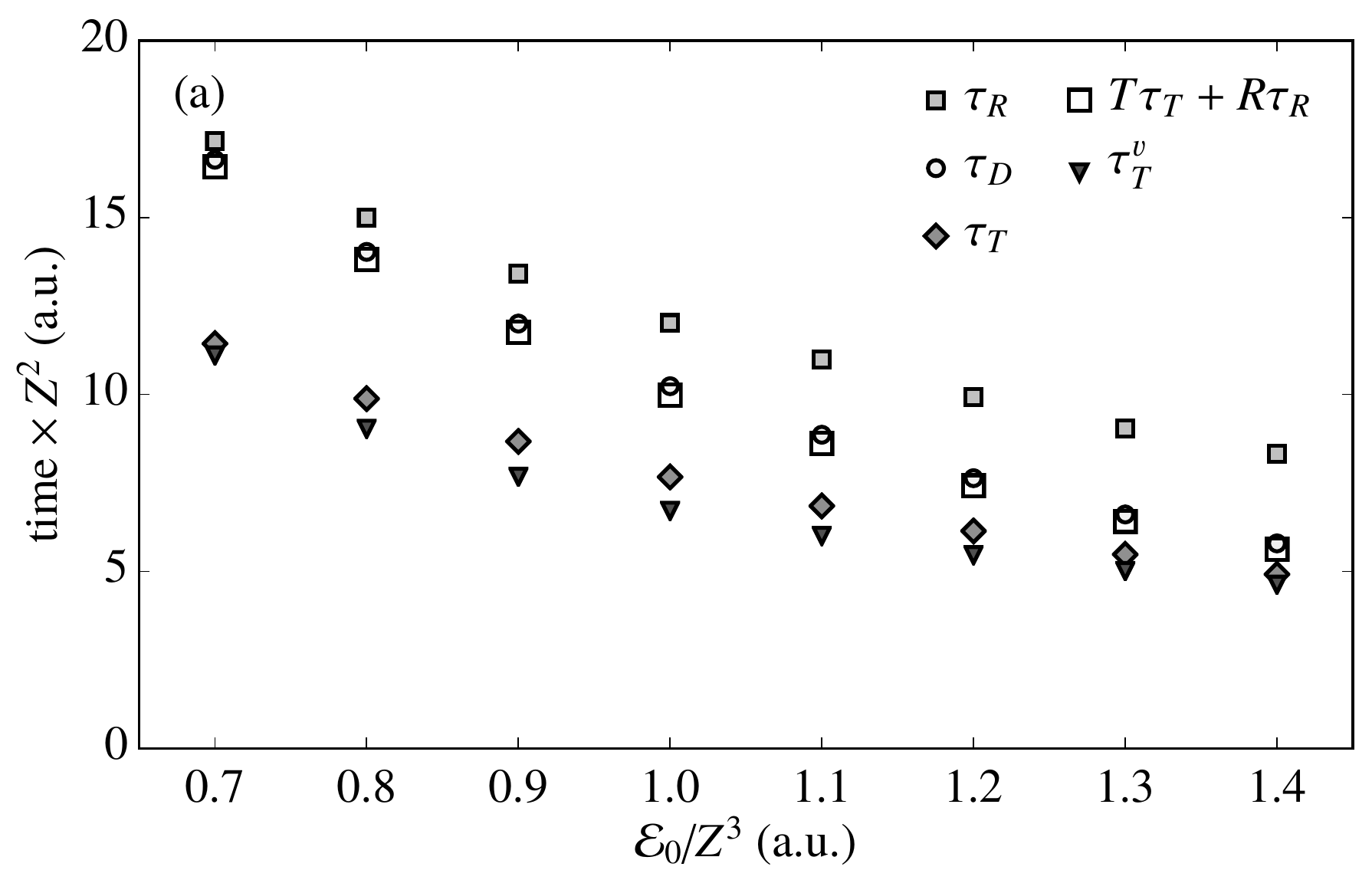}\\[2ex plus 2ex]
  \includegraphics[scale=0.45]{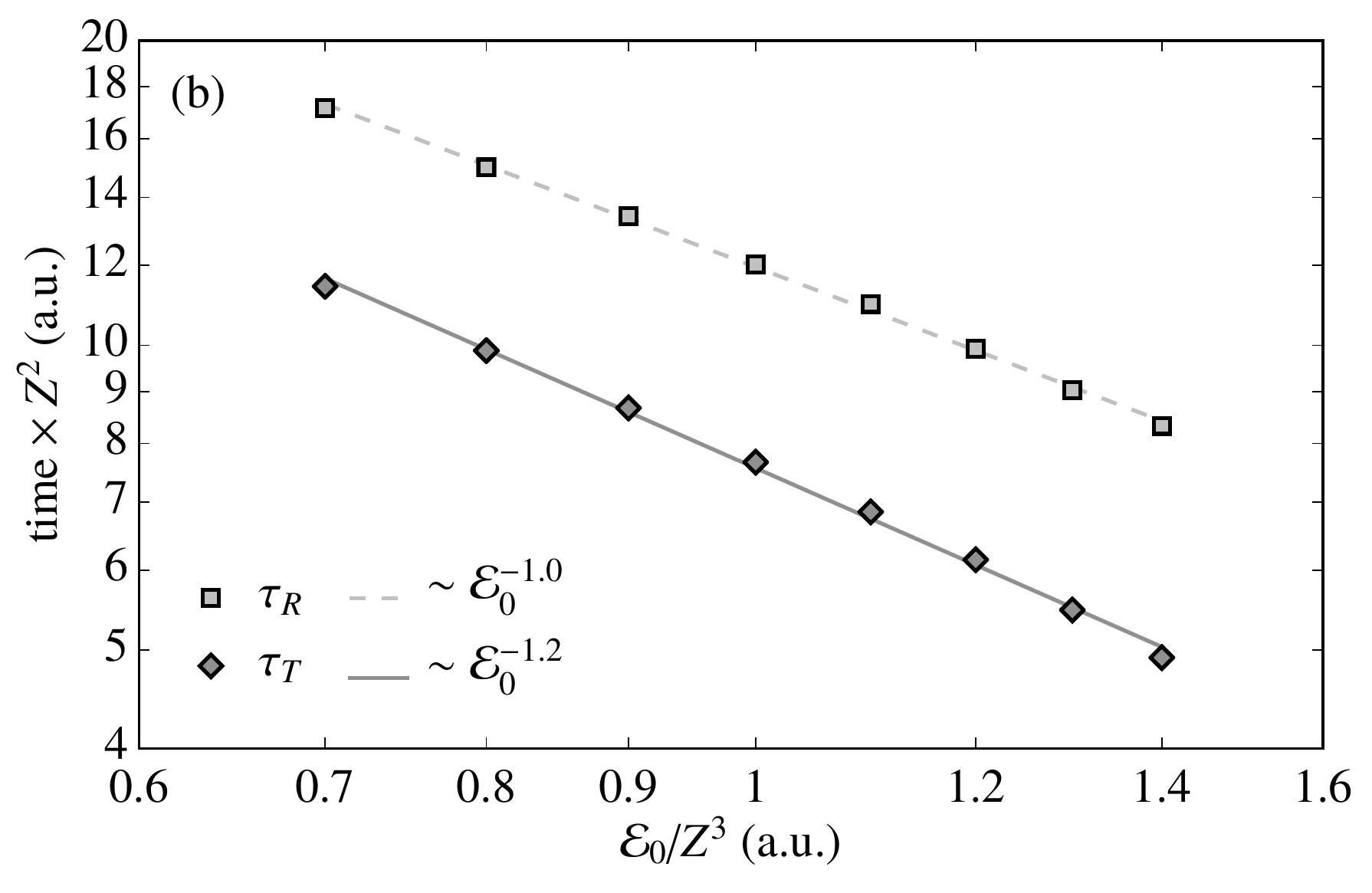}
  \caption{(a): Reflection time $\tau_R$, dwell time $\tau_D$, tunneling
    time $\tau_T$, the weighted sum $T\tau_T+R\tau_R$, and tunneling
    time $\tau_T^v$ for different electric field strengths
    $\mathcal{E}_0$ at a fixed Keldysh parameter $\gamma=0.25$.
    (b):~Reflection time $\tau_R$ and tunneling time of subfigure (a) on
    a double logarithmic scale.  For specific definitions of the various
    times see the main text.}
  \label{fig:delay_times}
\end{figure}

The two definitions of the dwell time \eqref{eq:t-dwell} and
\eqref{eq:tau_D}, respectively, provide us a valuable consistency check
of the quantum-clock approach.  To be consistent, the dwell time as
determined by the quantum clock \eqref{eq:tau_D} must agree with the
dwell time \eqref{eq:t-dwell}, which does not rely on the concept of a
quantum clock.  Our numerical results shown in Fig.~\ref{fig:dwell}
confirm this property.  For the chosen parameters, both dwell times
agree up to a discrepancy of about 4\,\%.  For larger field strengths
$\mathcal{E}_0$ this small discrepancy tends to be systematically larger
than for small field strengths.  This can be attributed to the fact that
the dwell time becomes small for large field strengths and therefore
the relative accuracy of the dwell-time determination is reduced. 

The tunneling time $\tau_T$, the reflection time $\tau_R$, as well as
the dwell time $\tau_D$ are presented in Fig.~\ref{fig:delay_times}(a).
Furthermore, the figure shows the weighted sum $T\tau_T+R\tau_R$, which
is close to the dwell time $\tau_D$.  But in contrast to the
relation~\eqref{eq:split_dwell_time}, $T\tau_T+R\tau_R$ does not
strictly agree with $\tau_D$ due to the nonlinearity of the
function~$\braket{\op{T}}_\mathrm{free}(t)$.  Because $\tau_D$ equals
approximately the weighted average $T\tau_T+R\tau_R$, the dwell time
$\tau_D$ lies always between the tunneling time $\tau_T$ and the
reflection time $\tau_R$.  For the in Fig.~\ref{fig:delay_times}
considered parameter range, the reflection time $\tau_R$ is larger than
the tunneling time $\tau_T$ by a factor of approximately~1.6.  This can
be intuitively understood as tunneling electrons enter the barrier and
then leave the barrier at the exit while reflected electrons move under
the barrier until they get reflected at the barrier exit and then travel
back into the direction of the atomic core.  The tunneling time as well
as the reflection time decrease with increasing electric field strength
$\mathcal{E}_0$.  In fact, we find a power-law behavior for $\tau_T$ as
well as for $\tau_R$, $\tau_T\sim \mathcal{E}_0^{-1.2}$ and
$\tau_R\sim \mathcal{E}_0^{-1}$ as shown in \ref{fig:delay_times}(b).
This scaling behavior is similar to the scaling of the Keldysh time
$\tau_K=\sqrt{-2E_Em}/(e\mathcal{E}_0)$, which has been identified as a
lower bound for the tunneling time in ionization by an instantaneously
turned on field
\cite{McDonald:Orlando:etal:2013:Tunnel_Ionization,Orlando:McDonald:etal:2014:Identification_of}.
In fact, the Keldysh time is always smaller than the tunneling time
$\tau_T$ by a factor of about four for the parameter range of
Fig.~\ref{fig:delay_times}.

\subsection{Relation to the virtual-detector approach}

In Ref.~\cite{Teeny:2016:virt_detect}, we studied tunneling times for the
same kind of system and the same parameter regime as in this article by
a different approach, the so-called virtual-detector method.  It is the
purpose of this section to relate the virtual-detector approach to the
quantum-clock approach.  We are going to demonstrate that both
complementary methods give compatible results for the tunneling time in
strong-field ionization.

\begin{figure}
  \centering
  \includegraphics[scale=0.45]{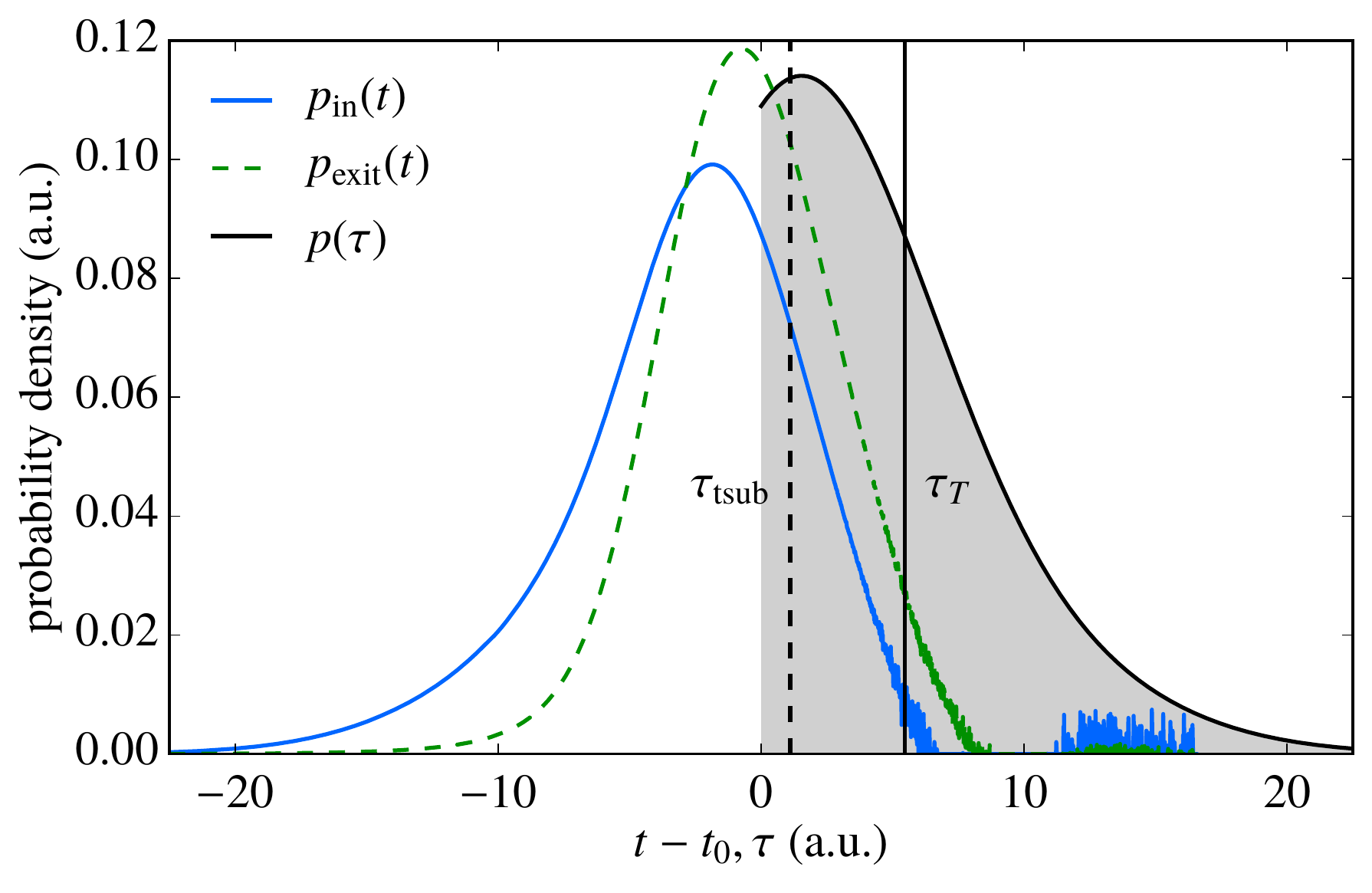}
  \caption{Probability distributions $p_{\mathrm{in}}(t)$ and
    $p_{\mathrm{exit}}(t)$ to enter or to leave the tunneling barrier
    into the direction away from the atomic core as functions of time
    and the probability distribution $p(\tau)$ of the tunneling time as
    derived from $p_{\mathrm{in}}(t)$ and $p_{\mathrm{exit}}(t)$ for
    $\mathcal{E}_0=1.2\,\mathrm{a.u.}$ and $Z=1$.  The vertical black
    dashed line indicates the time $\tau_\mathrm{tsub}$, which is the
    distance of the positions of the maxima of the distributions
    $p_{\mathrm{in}}(t)$ and $p_{\mathrm{exit}}(t)$.  The vertical black
    solid line corresponds to the tunneling time~$\tau_T$.}
  \label{fig:virtual_detector}
\end{figure}

The central idea of the virtual-detector approach is to determine the
electron's probability density flow at the entry line and the exit line
of the tunneling barrier at the parabolic coordinates
$\xi=\xi_\mathrm{in}$ and $\xi=\xi_\mathrm{exit}$ as functions of the
time $t$.  Integrating the probability density flow along these lines
gives the quantities $D\xi_\mathrm{in}(t)$ and $D\xi_\mathrm{exit}(t)$.
Furthermore,
$D_{\xi_\mathrm{in}}(t)\,\Theta(D_{\xi_\mathrm{in}}(t))\,\Delta t$ is
proportional to the probability that the electron crosses during the
short time interval $[t, t+\Delta t]$ the entry line of the tunneling
barrier into the direction away from the atomic core.  Here $\Theta(x)$
denotes the Heaviside step function.  Similarly,
$D_{\xi_\mathrm{exit}}(t)\,\Theta(D_{\xi_\mathrm{exit}}(t))\,\Delta t$
is proportional to the probability that the electron crosses the exit
line of the tunneling barrier into the direction away from the atomic
core.  It is convenient to introduce the normalization constants
$N_\mathrm{in}$ and $N_\mathrm{exit}$ and the distributions
$p_\mathrm{in}(t)$ and $p_\mathrm{exit}(t)$ such that\pagebreak[1]
\begin{subequations}
  \begin{align}
    p_\mathrm{in}(t) & =
    \frac{1}{N_\mathrm{in}} 
    D_{\xi_\mathrm{in}}(t)\,\Theta(D_{\xi_\mathrm{in}}(t))
    \,,\\
    p_\mathrm{exit}(t) & =
    \frac{1}{N_\mathrm{exit}} 
    D_{\xi_\mathrm{exit}}(t)\,\Theta(D_{\xi_\mathrm{exit}}(t))
    \,,
  \end{align}
\end{subequations}
and
\begin{subequations}
  \begin{align}
    \int_{-\infty}^\infty p_\mathrm{in}(t) \,\D t & = 1 \,, \\
    \int_{-\infty}^\infty p_\mathrm{exit}(t) \,\D t & = 1 \,.
  \end{align}
\end{subequations}
The functions $D_{\xi_\mathrm{in}}(t)$ and $D_{\xi_\mathrm{exit}}(t)$
have both a unique global maximum; see Fig.~\ref{fig:virtual_detector}.
In Ref.~\cite{Teeny:2016:virt_detect}, we defined the distance of the
positions of these maxima as the tunneling time $\tau_\mathrm{tsub}$,
which was also determined numerically for the same parameters as in this
article.

Comparing $\tau_\mathrm{tsub}$ and $\tau_T$, one finds that
$\tau_\mathrm{tsub}$ is smaller than $\tau_T$ by a factor of about
three.  This discrepancy is a direct consequence of the different
definitions of $\tau_\mathrm{tsub}$ and $\tau_T$.  It does not
necessarily indicate a conflict between the quantum-clock approach and
the virtual-detector approach.  As we will show in the following, the
discrepancy arises essentially because taking the difference between two
expectation values of two times is not equivalent to determining the
expectation value of a time delay.  Due to the approximate symmetry of
$p_{\mathrm{in}}(t)$ and $p_{\mathrm{exit}}(t)$ around their maxima,
$\tau_\mathrm{tsub}$ equals approximately the difference between the
expectation values for the moments of entering and leaving the tunneling
barrier.  The tunneling time $\tau_T$, however, is derived from an
expectation value of a clock operator which determines directly the time
spend in the tunneling barrier.  As a quantum mechanical observable, the
time that corresponds to this operator has some intrinsic distribution.
Neglecting possible quantum correlations between entering and leaving
the tunneling barrier, one can reconstruct this distribution from
$p_{\mathrm{in}}(t)$ and $p_{\mathrm{exit}}(t)$ by assuming that the
probability to spend the nonnegative time $\tau$ in the tunneling
barrier is proportional to the product of $p_{\mathrm{in}}(t)$ at the
entry time $t$ and $p_{\mathrm{exit}}(t+\tau)$.  Integrating over all
possible entry times yields then the probability distribution
\begin{equation}
  p(\tau) =
  \frac{1}{N}\int_{-\infty}^{\infty} 
  p_{\mathrm{in}}(t) p_{\mathrm{exit}}(t+\tau)\,\D t\,,
\end{equation}
where $N$ is a normalization constant such that
\begin{equation}
  \int_{0}^\infty p(\tau) \,\D \tau = 1 \,.
\end{equation}
Note that due to causality reasons, $p(\tau)$ vanishes for $\tau<0$.
The corresponding expectation value of the tunneling time of the
virtual-detector approach is then
\begin{equation}
  \tau_T^v = \int_{0}^\infty \tau p(\tau) \,\D \tau \,.
\end{equation}
For comparison with the tunneling time $\tau_T$ of the quantum-clock
approach, the time $\tau_T^v$ is also indicated in
Fig.~\ref{fig:delay_times}(a).  In contrast to tunneling delay
$\tau_\mathrm{tsub}$, $\tau_T^v$ is very close to the tunneling time
$\tau_T$ of the quantum-clock approach.  As one can see in
Fig.~\ref{fig:virtual_detector}, the distribution $p(\tau)$ has a
maximum approximately at $\tau_\mathrm{tsub}$.  Thus,
$\tau_\mathrm{tsub}$ corresponds to the most probable tunneling time.
Due to the distribution's asymmetry that results from the causality
condition $p(\tau)=0$ for $\tau<0$, the expectation value of $p(\tau)$
is shifted away from the position of the maximum to larger values, which
explains the \mbox{factor-3} discrepancy between $\tau_\mathrm{tsub}$
and $\tau_T$.

\vspace*{0pt plus 30pt}%

\section{Conclusions}
\label{sec:Conclusions}

We determined the tunneling time of strong-field tunnel ionization from
a two-dimensional Coulomb potential by utilizing the
Salecker-Wigner-Peres quantum clock.  A mean tunneling time of the order
of four times the Keldysh time was found by this method.  In
Refs.~\cite{McDonald:Orlando:etal:2013:Tunnel_Ionization,Orlando:McDonald:etal:2014:Identification_of},
the Keldysh time was identified as the time it takes the ground state in
the presence of a driving electric field to evolve into the quasistatic
resonance state after an instantaneous turn-on of the field.  Thus, the
obtained tunneling time may be seen as an indication that the Keldysh
time characterizes the time to develop quasistatic resonance state also
in the case of a continuously evolving electric field.

The quantum-clock approach is complementary to the virtual-detector
approach that was taken in an earlier work. The mean tunneling time as
identified by the quantum clock could be reproduced from the probability
distributions for the moments of entering and leaving the tunneling
barrier, which were determined by the virtual-detector approach, and
assuming statistical independence of entering and leaving the tunneling
barrier.  This has two implications.  The fact that we find almost the
same mean tunneling time by two very different theoretical approaches is
a strong indication that in agreement with there \emph{is} a nonzero
tunneling time and tunneling is neither instantaneous nor superluminal,
in agreement with experimental
findings~\cite{Landsman:Weger:etal:2014:Ultrafast_resolution} and other
theoretical findings, e.\,g., in
Ref.~\cite{Zimmermann:Mishra:etal:2016:Tunneling_Time}.  Furthermore,
the agreement of both methods may be interpreted as that entering and
leaving the tunneling barrier are nearly statistically independent, up
to the causality condition that the electron cannot leave the tunneling
barrier before it has entered the barrier.

A substantial difference was found between the most probable tunneling
time and the mean tunneling time, which is larger than the former.  This
may be of experimental relevance.  An ideal series of experimental
measurements should yield the mean tunneling time.  As pointed out in
Ref.~\cite{Zimmermann:Mishra:etal:2016:Tunneling_Time}, only a
post-selected subset of ionized electrons may be actually detected,
corresponding to, e.\,g., electrons with the most probable momentum or
with the most probable tunneling time.  In the latter case, an
experiment would find the most probable tunneling time, not the mean
tunneling time.  Consequently, it is essential to distinguish between
the mean tunneling time and other kinds of tunneling time when comparing
tunneling times quantitatively.

\vspace*{10pt plus 30pt}%
\appendix*

\section{Numerical methods}
\label{sec:methods}

Coupling the quantum clock Hamiltonian to the electron's Schr{\"o}dinger
Hamiltonian yields an Hamiltonian for a \mbox{$N$-component} wave
function.  Because the clock Hamiltonian is diagonal, however, the
resulting equation of motion for this \mbox{$N$-component} wave function
separates into $N$ independent one-component Schr{\"o}dinger equations which
are given by the usual Schr{\"o}dinger equation for the electron modified by
an additional clock potential.  At least in principle, these
time-dependent Schr{\"o}dinger equations may be solved numerically by any
standard method.

Due to the Coulomb potential's singularity and the weakness of the clock
potential (compared to the Coulomb potential) a naive application of
some standard method is likely to fail.  Numerical difficulties in the
application of the quantum-clock approach may be circumvented by
employing a Trotter-Suzuki splitting scheme
\cite{Suzuki:1976:Generalized_Trotters}.  In this scheme, the time
evolution operator of the whole quantum system is split into products of
time evolution operators for the clock Hamiltonian and the Schr{\"o}dinger
Hamiltonian with the Coulomb potential plus the external electric field.
The clock Hamiltonian can be propagated exactly.  The remaining
Schr{\"o}dinger Hamiltonian is propagated by employing a Lanczos propagator
\cite{Park:Light:1986:Unitary_quantum,Beerwerth:Bauke:2015:Krylov_subspace}
and a fourth-order finite differences scheme for the discretization of
the Hamilton operator.

\bibliography{quantum-clock}

\end{document}